\documentclass{article}
\usepackage{arxiv}

\usepackage[numbers,sort&compress]{natbib}
\usepackage{fontawesome}
\usepackage{placeins}

\usepackage{xcolor}
\definecolor{blue}{rgb}{0.0, 0.0, 1.0}
	
\usepackage[utf8]{inputenc} 
\usepackage[T1]{fontenc}    
\usepackage{hyperref}       
\usepackage{url}            
\usepackage{booktabs}       
\usepackage{amsfonts}       
\usepackage{nicefrac}       
\usepackage{microtype}      
\usepackage{lipsum}		    
\usepackage{graphicx}
\usepackage{natbib}
\usepackage{doi}
\usepackage{pifont}         

\usepackage{wrapfig}
\usepackage{graphicx}

\newcommand{\ie}{\textit{i.e.,}~}
\newcommand{\eg}{\textit{e.g.,}~}
\newcommand{\resp}{\textit{resp.}~}
\newcommand{\etc}{\textit{etc.}}
\newcommand{\cf}{\textit{cf.}~}

\title{Amman City, Jordan: Toward a Sustainable City from the Ground Up}

\author{ \href{https://orcid.org/0000-0002-9559-2293}{\includegraphics[scale=0.06]{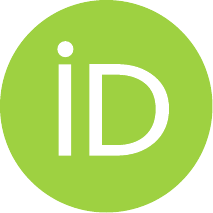}\hspace{1mm}Ra'Fat A.~Al-Msie'deen}\thanks{https://rafat66.github.io/Al-Msie-Deen/} \\
	Department of Software Engineering\\
	Faculty of IT, Mutah University\\
	Mutah 61710, Karak, Jordan \\
	\texttt{rafatalmsiedeen@mutah.edu.jo} \\
}

\renewcommand{\shorttitle}{\textit{Ra'Fat Al-Msie'deen}: Amman City, Jordan: Toward a Sustainable City from the Ground Up.}

\hypersetup{
pdftitle={Amman City, Jordan: Toward a Sustainable City from the Ground Up},
pdfsubject={Smart City},
pdfauthor={Ra'Fat Al-Msie'deen},
pdfkeywords={Amman City, Smart City (SC), Information and Communication Technologies (ICT), Internet of Things (IoT), Cloud Computing (CC), Artificial intelligence (AI), Big Data Analytics (BDA)},
}

\begin{document}
\maketitle

\begin{abstract}
	The idea of smart cities (SCs) has gained substantial attention in recent years. The SC paradigm aims to improve citizens' quality of life and protect the city's environment. As we enter the age of next-generation SCs, it is important to explore all relevant aspects of the SC paradigm. In recent years, the advancement of Information and Communication Technologies (ICT) has produced a trend of supporting daily objects with smartness, targeting to make human life easier and more comfortable. The paradigm of SCs appears as a response to the purpose of building the city of the future with advanced features. SCs still face many challenges in their implementation, but increasingly more studies regarding SCs are implemented. Nowadays, different cities are employing SC features to enhance services or the residents quality of life. This work provides readers with useful and important information about Amman Smart City.
\end{abstract}


\section{Introduction: smart cities} \label{sec:In}

Along with the evolution of society, there has been a significant development of ICT. Technologies like the Internet of Things (IoT) and cloud computing have gained importance in recent times \cite{BelliniPierfrancesco}. IoT is the heart of SCs. These technologies are becoming a main part of the evolution of cities. Therefore, urbanization and ICT have led to the emergence of the SC paradigm \cite{Sanchez-Corcuera19}.
SCs have been greatly developed and have significantly extended their capabilities in recent years.
In general, SC is an urban area that exploits ICT to improve the quality of life of its citizens (\ie the heart of a SC) as well as the sustainability and efficiency of its operations.	

SCs are those modern cities that employ modern ICT facilities to satisfy their citizens \cite{BushraHamid}. According to recent research studies, by 2030, sixty percent of the country’s population will be occupied in modern cities \cite{AlDairiAnwaar}. Therefore, living in modern cities poses a great challenge for governments, as it is not easy to manage these large cities using traditional methods, especially when the number of citizens increases significantly and thus the demand for services increases. Thus, modern SCs will be able to accommodate these large numbers of citizens and manage their various needs, as SCs are equipped with modern ICT facilities.

IoT technology is centered on embedding sensors into everyday objects and utilizing connectivity to enable information exchange for various applications. There are more objects available than people in a city, so the amount of connectivity that IoT devices hold is vast. Unlike the internet, which is a grid of networks, the IoT is a network of interconnected devices. IoT devices are crucial in people's daily lives in the city, managing large volumes of data. IoT devices can be viewed as a mesh of communicated devices that implies sending and operating devices that offer the capacity for information exchange among distinct platforms in a city \cite{JieTan}. IoT devices have lots of valuable applications when they appear in SCs \cite{KimTai}. A SC can be recognized as a city that is supplied with technology, like sensors and cameras that gather data; it is exploited to make critical decisions in the management of city operations \cite{RobinChataut}.

\begin{figure*}[!htb]
	\centering
	\includegraphics[width=\textwidth]{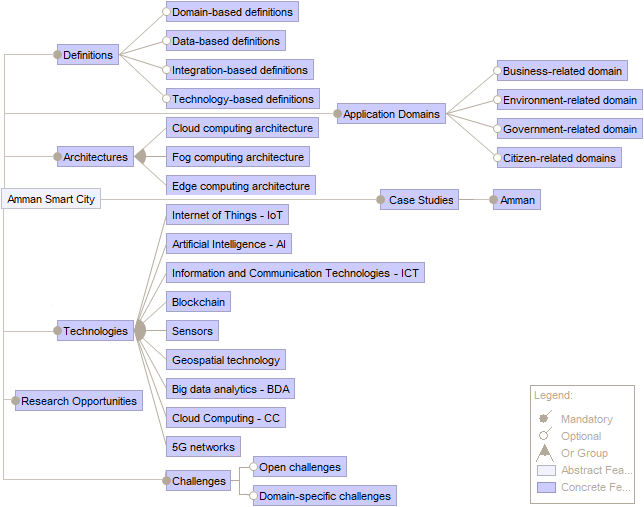}
	\caption{Main topics related to Amman Smart City.} \label{fig:SmartCityFM}
\end{figure*}

This article covers the main topics relevant to Amman smart city; mainly, it covers the following topics: smart city definitions, architectures, technologies, application domains, case studies, challenges, and research opportunities. Figure \ref{fig:SmartCityFM} summarizes the main topics of this article. This figure employs a feature model (FM) to discover the topics covered in this study \cite{DBLPMsiedeenSH123} \cite{DBLPAlMsiede456}. Interested readers can find more information about this FM in \cite{DBLPAlMsiedeenHSUV} and \cite{DBLPphdAlMsieDeen}.

In the last two decades, the concept of SC has become more and more popular, but there is still confusion around what a SC is, specifically since many similar concepts are often utilized interchangeably \cite{AlbinoVito}. Nowadays, SC paradigm is a promising and applicable approach. Every SC has characteristics, requirements, and infrastructure, and people expect a lot from this city so that it provides them with flexible, comfortable, and high-quality services. In this section, the author defines the concept of a smart city based on his knowledge. Table \ref{tab:SCDefinitions} presents the author's definition of a smart city.

\begin{table*}[!htb]
	\begin{center}
		\caption{Definition of smart city: the author's definition.}
		\begin{tabular}{|l|p{6cm}|l|} \hline 
			\textbf{Reference}                       &  \textbf{Definition}   & \textbf{Category} \\\hline			
			Al-Msie'deen (author's definition) & A smart city is a city that is prepared based on up-to-date traditional (streets, hospitals, and schools) and modern communication (ICT, AI, and IoT) infrastructure to improve the quality of life for its citizens (city services such as education, healthcare, traffic, and so on). & Technology-based definitions \cite{Palmisano} \cite{HarrisonEHHKPW10} \cite{SuKehua} \cite{CoeAmanda} \\\hline
		\end{tabular}
		
		\label{tab:SCDefinitions}
	\end{center}	
\end{table*}

Many definitions of SCs exist (the concept was first utilized in the 1990s) \cite{AlbinoVito}. A smart city, according to the author, is a city that is prepared based on up-to-date traditional (streets, hospitals, and schools) and modern communication (ICT, AI, and IoT) infrastructure to improve the quality of life for its citizens (city services such as education, healthcare, traffic, and so on).

Since the SC paradigm was created, several researchers have tried to present the most suitable architecture according to ICT solutions \cite{Sanchez-Corcuera19}. Because of the diverse requirements and everyday circumstances in cities that have executed SC architectures, these architecture executions have not been following a standard, and consequently, they have diverse characteristics \cite{LongoFrancesco}. In fact, there isn't an ideal architecture for SCs.

Cloud computing architecture is split into numerous layers (\eg infrastructure-as-a-service). Each of the layers consumes the services offered by the other layers and also offers its own services to the other layers.
Fog computing architecture is an extension of cloud computing architecture. This extension is achieved by enhancing the performance and duties of the network's end nodes. In this type of architecture, the data storage and processing devices are installed on the same local network. Thus, the fog computing architecture creates low-latency communication between its components. This architecture is utilized for applications where low latency is needed.
Whereas edge computing follows the assumption that if the data is produced at the network's edge (\eg sensor), then it would be more effective to process that data at the same edge \cite{Sanchez-Corcuera19}.
These architectures (\ie cloud, fog, and edge computing architectures) are complementary and can be executed in the same city, where each architecture has its own characteristics \cite{AitorAlmeida}.

Nowadays, technologies are the backbone of several SCs around the world. The use of the latest technologies (\eg IoT, big data, or cloud computing) inside the city aims at solving city-critical problems and improving a variety of public services and functions within the city. This section presents the major technologies that define the smartness of a city. Table \ref{tab:SmartTechnologies} presents the most important technologies for SC.

\begin{table*}[!htb]
	\begin{center}
		\caption{SC technologies: major technologies that define the smartness of a city.}
		\begin{tabular}{|l|p{12.6cm}|l|} \hline 
			\textbf{Technology}                       &  \textbf{Definition/Description}    & \textbf{Reference} \\\hline
			
			Internet of Things (IoT) & "The network of devices that contain the hardware, software, firmware, and actuators which allow the devices to connect, interact, and freely exchange data and information." &  \cite{RossRon} \\
			& "As used in this publication, user or industrial devices that are connected to the internet. IoT devices include sensors, controllers, and household appliances." & \cite{AkramMehwish} \\\hline
			
			ICT & "Includes all categories of ubiquitous technology used for gathering, storing, transmitting, retrieving, or processing information (\eg microelectronics, printed circuit boards, computing systems, software, signal processors, mobile telephony, satellite communications, networks). ICT is not limited to information technology but reflects the merging of information technology and communications." & \cite{DOD520044} \\
			& "ICT encompasses all technologies for the capture, storage, retrieval, processing, display, representation, organization, management, security, transfer, and interchange of data and information." & \cite{HoganMichael} \\\hline
			
			Artificial Intelligence (AI)& AI is a technology that assists computers and other smart devices (\eg mobile devices, smart sensors, and wearable devices) to simulate human intelligence in addition to problem-solving capabilities. The two most common sub-disciplines of AI are Machine Learning (ML) and Deep Learning (DL). & \cite{IBMAI}  \\\hline
			
			Sensors & "A device that produces a voltage or current output that is representative of some physical property being measured (\eg speed, temperature, and flow)." &  \cite{StoufferKeith} \\
			& "A portion of an IoT device capable of providing an observation of an aspect of the physical world in the form of measurement data." & \cite{FaganMichael} \\\hline
			
			Big data analytics & "Big data analytics refers to those advanced technologies that are involved in analyzing large-scale heterogeneous datasets, big data mining, and statistical analysis." & \cite{PramanikMd} \\
			&"Big data analytics is a phenomenon that analyses large volumes of data using sophisticated tools and techniques to extract valuable insights and to solve business use-cases." & \cite{ZhangJustin} \\\hline
			
			5G networks & "The term 5G stands for the fifth generation of network technology. The 5G network was deployed worldwide in 2019 instead of the 4G network that was introduced in 2009. Inside SCs, the 5G network connects approximately everyone and everything, involving machines, objects, and tools. The main features of 5G networks are large bandwidth, high speeds, lower latency times, and minor energy consumption." & \cite{SatheeshkumarR}, \cite{CasiniMarco}\\\hline
			
		\end{tabular}
		\label{tab:SmartTechnologies}
	\end{center}
\end{table*}

SC leverages modern technologies such as IoT, smart sensors, 5G networks, smart traffic lights, autonomous vehicles, Global Positioning System (GPS), and Electric Vehicle (EV) charging stations to improve transportation. Also, SC employs smart meters, smart grids, and renewable energy sources for effective energy management. Moreover, the environmental monitoring of SC is improved through air quality sensors and smart waste bins. Furthermore, the public safety sector in SC benefits from smart surveillance cameras and emergency response systems.

Numerous sectors inside SC benefit from modern technologies. The healthcare sector exploits telemedicine platforms and wearable health devices to improve its services. Furthermore, citizen engagement is facilitated through mobile apps and online participation platforms. Additionally, smart buildings use building management systems and smart security systems. Finally, data analytics and AI underpin these technologies for constructing more effective, sustainable, and livable urban environments.

A SC encompasses numerous domains where the advent of ICT leads to pivotal transformation and change. In this work, the author decided to adopt Yin's taxonomy for application domains and divide domains into four categories (\ie business, citizen, environment, and government-related categories) with a broader extension \cite{YinXCWCD15}.

The remaining sections of this article cover the following: Section \ref{sec:Amman} presents a summary of Amman City as a sustainable city, and, finally, section \ref{sec:Summary} gives a summary of this work.

\section{Amman smart city} \label{sec:Amman}

Through the last decade, several cities around the world have attempted to change from a conventional city to a SC. But, in several cases, those endeavors have been unsuccessful, even though there have been considerable investments from the public and private sectors. Table \ref{tab:SmartTop2024} presents the top 10 smart cities in 2024 \cite{IMD}. 

The information presented in Table \ref{tab:SmartTop2024} is taken from the IMD smart city index for 2024. This index ranks 142 smart cities worldwide according to the data analyzed by scholars as well as the survey answers of 120 citizens in each city. Also, it provides an overview of how a city's infrastructure and technology influence its overall performance and the quality of life for its residents.

\begin{table*}[!htb]
	\begin{center}
		\caption{The top smart cities in 2024 \cite{IMD}.}
		\begin{tabular}{|p{4.9cm}|p{4cm}|c|c|p{3.9cm}|} \hline 
			\textbf{City}       &  \textbf{Country}  & \textbf{SC rank 2024} & \textbf{SC rank 2023} & \textbf{Change} \\\hline
			
			Zurich     &  Switzerland          & 1  & 1  & --- \\\hline			
			Oslo       &  Norway               & 2  & 2  & --- \\\hline
			Canberra   &  Australia            & 3  & 3  & --- \\\hline
			Geneva     &  Switzerland          & 4  & 9  & +5 $\bigtriangleup$ \\\hline
			Singapore  &  Singapore            & 5  & 7  & +2 $\bigtriangleup$ \\\hline
			Copenhagen &  Denmark              & 6  & 4  & -2 $\bigtriangledown$ \\\hline
			Lausanne   &  Switzerland          & 7  & 5  & -2 $\bigtriangledown$ \\\hline
			London     &  United Kingdom       & 8  & 6  & -2 $\bigtriangledown$ \\\hline
			Helsinki   &  Finland              & 9  & 8  & -1 $\bigtriangledown$ \\\hline
			Abu Dhabi  &  United Arab Emirates & 10 & 13 & +3 $\bigtriangleup$ \\\hline
			
			\multicolumn{5}{l}{Rankings out of 142 cities. $\bigtriangleup$ $\bigtriangledown$ Change from previous year. No arrow indicates no change.
			}\\
			
		\end{tabular}
		\label{tab:SmartTop2024}
	\end{center}
\end{table*}

This section presents Amman as a SC (\cf summary in Table \ref{tab:AmmanSC}). Amman ranks 128\textsuperscript{th} in the IMD SC Index 2024 (\ie 128 out of 142), while the city is ranked 135\textsuperscript{th} in the same index in 2023. Amman is the capital of Jordan. It is located in the northwest of Jordan. Amman has an area of about 7,579 km\textsuperscript{2}. The population of Amman is expected to reach six million in 2025. Amman's vision is to improve the quality of life for Amman's citizens with the utilization of practical technology and achieve resilient and sustainable development. The basic elements of the approach toward SC are automation using new technology, business process re-engineering, and the private sector. The city approach aims at identifying the essential resources to achieve the SC vision, assessing stakeholder concerns and user needs, and defining the potential projects to be included in the city roadmap.

\begin{table*}[!htb]
		\caption{Summary of Amman, analyzing if Amman has SC vision, platform, a roadmap designed, a particular department customized entirely to the evolution of Amman as SC, the SC domains the city covers, the projects that have been executed in Amman, and the used technologies.}
		\scalebox{0.907}{
		\begin{tabular}{|c|c|c|c|c|c|c|c|c|} \hline 
			\textbf{City name}       &  \textbf{Country}  & \textbf{SC rank 2024} & \textbf{SC rank 2023}  & \textbf{Change} & \textbf{Vision} & \textbf{SC platform} & \textbf{Roadmap designed} & \textbf{SC department} \\\hline
			
			Amman       &  Jordan  & 128 & 135  & +7 $\bigtriangleup$ & \ding{51} & \ding{51} & Yes &  No \\\hline
				
			\multicolumn{9}{l}{}\\\hline
			\multicolumn{5}{|l|}{}& Business & Citizen & Environment & Government \\\hline
			\multicolumn{5}{|l|}{\textbf{Domains}} & \ding{53} & \ding{53} & \ding{53} & \ding{53} \\\hline
			
			\multicolumn{9}{l}{}\\\hline
			\multicolumn{5}{|l|}{\textbf{Projects}}& Business & Citizen & Environment & Government \\\hline
			\multicolumn{5}{|l|}{E-payments} & \ding{51} & & & \\\hline
			\multicolumn{5}{|l|}{Smart traffic control system} & & \ding{51} & & \\\hline
			\multicolumn{5}{|l|}{Smart home system} & & & \ding{51} & \\\hline
			\multicolumn{5}{|l|}{Traffic monitoring platform} &&&& \ding{51} \\\hline
			\multicolumn{5}{|l|}{Flood Early Warning System (FEWS)} &&&&  \ding{51} \\\hline
			\multicolumn{5}{|l|}{Smart streetlight project} & & & \ding{51} & \\\hline
			\multicolumn{5}{|l|}{Expanding ride-sharing services}& & \ding{51} & & \\\hline
			\multicolumn{5}{|l|}{Parking and curbside management}& & \ding{51} & & \\\hline
			\multicolumn{5}{|l|}{Transportation electrification project}& & \ding{51} & & \\\hline
			\multicolumn{5}{|l|}{Smart road asset management project}& &  & & \ding{51} \\\hline
			\multicolumn{5}{|l|}{Dedicated walking and biking paths}& & \ding{51} & & \\\hline
			\multicolumn{5}{|l|}{The Bus Rapid Transit (BRT) project}& & \ding{51} & & \ding{51} \\\hline
			\multicolumn{5}{|l|}{Expanded BRT feeder services by neighborhood}& & \ding{51} & & \ding{51} \\\hline
			\multicolumn{5}{|l|}{Areas guide (Amman explorer)} & & \ding{51} &  & \\\hline
			\multicolumn{5}{|l|}{Digital marketing and e-commerce} & \ding{51} &  &  & \\\hline
			\multicolumn{5}{|l|}{Smart energy meters for smart grids} &  &  & \ding{51} & \\\hline
			\multicolumn{5}{|l|}{Waste management (waste sorting and recycling)} &  &  & \ding{51} & \\\hline
			
			\multicolumn{9}{l}{}\\\hline
			& IoT & ICT & Blockchain & AI & CC \& BDA & Geospatial & Sensors \& Cameras & 5G networks \\\hline
			\textbf{Technologies} & \ding{53} & \ding{53} & \ding{53} & \ding{53} & \ding{53} & \ding{53} & \ding{53} & \ding{53}  \\\hline
				
		\end{tabular}}
		\label{tab:AmmanSC}
\end{table*}

Amman was built as a traditional city with basic infrastructure (\ie buildings, streets, hospitals, universities, schools, and so on). Currently, the city adopts modern technology infrastructure, such as 4G/5G networks, high-speed communication lines, machines and computers with high specifications, high-speed internal networks, data centers of high standards, disaster recovery methodologies, secure firewalls, blockchain, cybersecurity, AI, \etc

The digital transformation in Amman led to several improvements in various domains of the city. Some of these improvements include: 100\% of e-services completed; 134 services available online; paperless procedures at Greater Amman Municipality (GAM); the GAM app; the Unified Inspection System (UIS); QR code apps; and a decrease in the number of visitors to city government facilities by 98\%.

In 2019, GAM developed a SC roadmap for Amman \cite{GAM}. This roadmap exploits the latest improvements in technologies (AI and big data) to resolve the city’s urban planning requests. Moreover, this roadmap tries to increase the use of e-services and automate procedures in several ministries to attain transparency and increase the efficiency of government services and functions. Also, the roadmap intends to improve city services and reduce traffic congestion inside Amman.

Many benefits have been achieved in the city of Amman after applying the SC paradigm. Some of these benefits include e-payments (\eg eFAWATEERcom), end-to-end services, efficiency of city functions and services (cost, quality, and speed), simple procedures (\ie e-transactions), traffic congestion management, paperless procedures (\ie eco-friendly), employee training, \etc

\begin{figure*}[!htb]
	\centering
	\includegraphics[width=\textwidth]{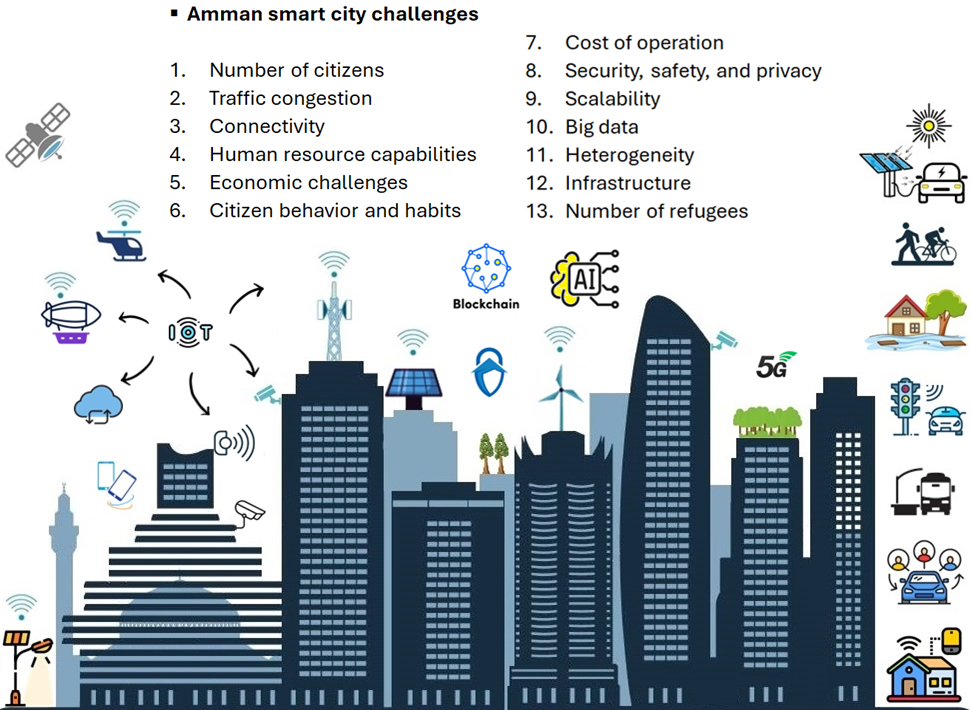}
	\caption{Amman smart city challenges.} \label{fig:Amman}
\end{figure*}

Amman, a highly urbanized capital, faces several open challenges. Some of those challenges are rapidly growing population and traffic congestion, large number of refugees in Jordan, human resources capabilities for smart solutions, economic challenges globally and regionally, citizen behavior and habits, and so on (\cf Figure \ref{fig:Amman}).

As some researchers center their study on a specific domain of SCs, Table \ref{tab:specificChallenges} presents the challenges of specific domains inside SCs, such as agriculture, healthcare, and so forth.

\begin{table}[h!]
	\begin{center}
		\caption{Domain specific challenges for SCs.}
		\begin{tabular}{|l|p{12.5cm}|l|} \hline 
			\textbf{Domain}             &  \textbf{Challenge(s)}  & \textbf{References} \\\hline
			
			Innovation         &  Strategic vision (\eg the ambiguity of being a SC); organizational capabilities and agility (\eg administrative constraints and human resources); technology domestication (\eg technological determinism); ecosystem development (\eg governance models and business models); and transboundary innovation (\eg scalable boundary-spanning collaborations).   &  \cite{RuiJosé} \\\hline
			
			IoT & Technical challenges (\eg interoperability, standardization, data privacy/security, and network/device compatibility); social challenges (\eg accessibility, inclusiveness, user acceptance/adoption, and ethical/legal considerations); and economic challenges (\eg return on investment, implementation/deployment processes, and cost/funding).    &  \cite{LiuJie} \\
			& Challenges related to wearable and autonomous computing systems. & \cite{DomenicoBalsamo} \\
			& Challenges related to security and privacy, smart sensors, networking, and big data analytics. & \cite{ShahSyed} \\\hline
			
			Artificial intelligence (AI)    & Challenges related to security/surveillance; energy consumption/distribution; operationalizing new technologies; predicting future needs; AI’s ethical use; and measuring the environmental effects of AI infrastructure. & \cite{AndrasSzorenyi} \\\hline
			
			Agriculture &   Challenges related to the smart farm, such as the high cost of technology, constructing farms inside cities, the energy outlay, the high real estate cost, the need for IT skills, and a lack of awareness.  &  \cite{AlirezaMoghayedi} \\\hline
			
			Waste management          &  Lack of strict government regulatory policies, proper financial planning, and benchmarking processes.   &  \cite{YadavHoney} \\
			& Electronic waste (e-waste) and the identification of waste materials before the separation process (an expensive job). & \cite{DanutaSzpilko} \\\hline
			
			Traffic             &  Traffic congestion, heterogeneous traffic flow, capacity of the roads, vehicle-to-vehicle communication, and berth scheduling.   & \cite{AuwalMusa} \\\hline
			
			Energy              &  Implementation challenges (\eg technical complexity, resistance to change, privacy and security concerns, high implementation costs, and limited data access); variability (\resp intermittency) of renewable energy sources; and managing energy consumption in cities (\ie mismatch between energy supply and demand).   & \cite{PandiyanPitchai}  \\\hline
			
			Healthcare & Technological challenges (\eg costs, poor data quality, and data governance/ethical concerns); organizational challenges (\eg poor planning, poor leadership, and change culture); environment challenges (\eg poor computer skills and end user behavior). & \cite{RenukappaSuresh} \\
			& Challenges related to health monitoring. & \cite{MshaliHaider} \\\hline
			&     &  \\\hline

		\end{tabular}
		\label{tab:specificChallenges}
	\end{center}
\end{table}

The multidisciplinary nature of SCs creates multiple research challenges. Table \ref{tab:openChallenges} presents the open challenges of SC (\ie author's opinion).

\begin{table*}[!htb]
	\begin{center}
		\caption{Open challenges for SCs: author's opinion.}
		\begin{tabular}{|l|p{13.6cm}|l|} \hline 
			\textbf{Reference}               &  \textbf{Challenges}    \\\hline
			
			Al-Msie'deen (author's opinion) & City's infrastructure (\ie traditional, ICT, and IoT infrastructure); data security, safety, and privacy; interoperability; governance and policy issues; coordination and collaboration between private and public sectors; investment and funding; mobility; technological integration; connectivity; public awareness and acceptance; and sustainable energy. \\\hline
			
		\end{tabular}
		\label{tab:openChallenges}
	\end{center}
\end{table*}

The ICT sector evolves rapidly in Jordan, especially in Amman. The ICT sector plays an important role in Gross Domestic Product (GDP), where it is considered the third biggest contributor to GDP. Thus, one of the main drivers of Amman’s sustainable growth is the ICT sector. This sector can assist in modernizing the public sector and providing smart infrastructure, such as smart building and transport, e-services, and data collection and analytics. Finally, this sector can also offer support for evidence-based policymaking.

\begin{figure*}[!htb]
	\centering
	\includegraphics[width=\textwidth]{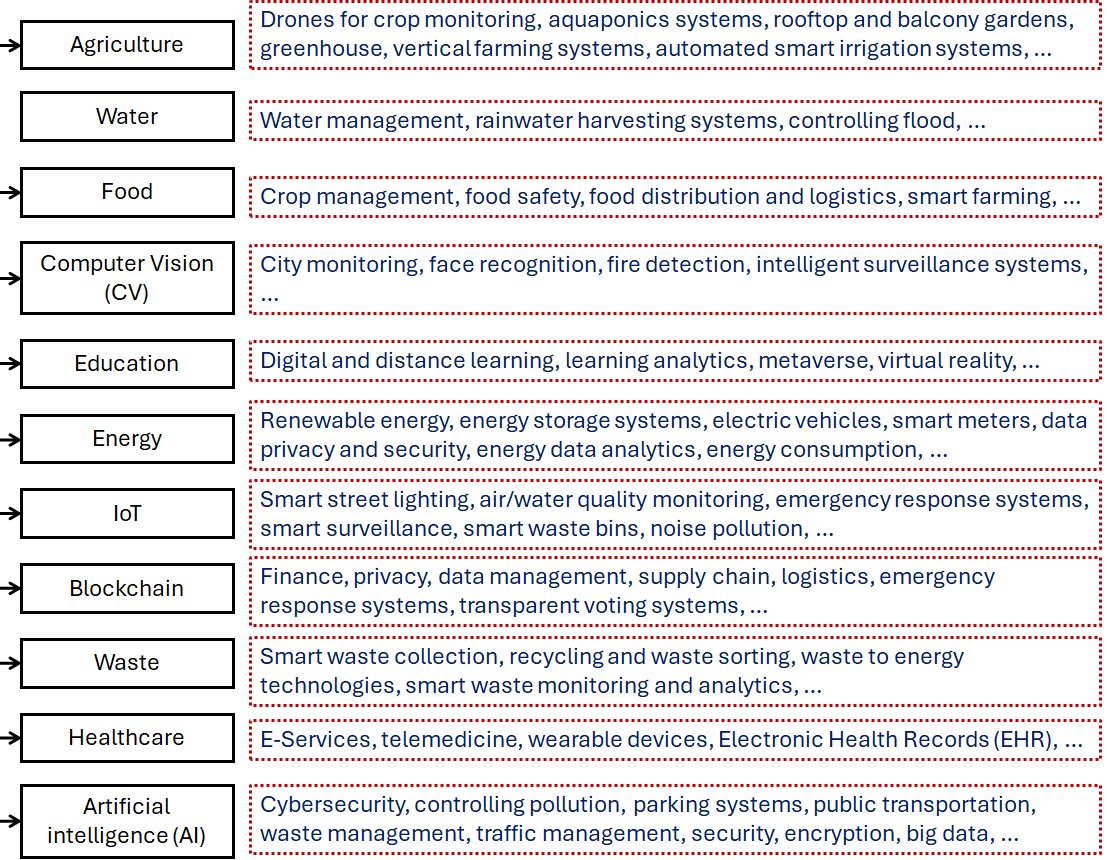}
	\caption{Smart cities research opportunities.} \label{fig:researchOpportunities}
\end{figure*}

Based on the knowledge acquired through the literature review, Figure \ref{fig:researchOpportunities} presents the research opportunities that are promising and related to the SC paradigm. These research opportunities are also applicable in the city of Amman.

Several studies have been done in the fields of water and food management inside SCs. These fields have not been very well explored yet; thus, it is possible to advance these fields in several areas, such as crop management, transgenic research, water management, water storage, and so on.

Computer Vision (CV) is a multidiscipline field including engineering, biology, mathematics, psychology, and physics disciplines. It aims to address how computers obtain useful interpretations of digital images and videos. Thus, it improves a computer's ability to obtain, process, analyze, and interpret digital media. CV involves vast subdomains, such as object tracking, motion estimation, virtual reality, and face recognition \cite{ZouLan}. CV studies are quite important to SCs, especially in the healthcare domain. Also, it can be utilized in other domains, like city monitoring, advertisement, traffic management systems, and logistics \cite{Sanchez-Corcuera19}. Pushing this field forward and utilizing modern techniques inside SCs might be a key advancement.

In the upcoming years, the SC paradigm in healthcare is expected to become more significant due to the growing urban population and rising healthcare demands. This will present new opportunities for both healthcare providers and patients \cite{MohammadzadehZahra}. Recently, the most research effort has been done in the healthcare area inside SCs. Several improvements were made regarding this domain in bioinformatics, biomedical image analysis, and so on. In fact, more studies are needed to improve this domain in order to be totally accepted by patients in the city. Also, more research is needed to improve electronic services for both clinics and hospitals.

Regarding the education sector, there are numerous studies that have been done in the literature \cite{AfzalBadshah} \cite{KostadinovaDicheva} \cite{ShubhamDubey}. Smart education is the main component of SC development. SCs always include new education services and applications for citizens to exploit. For me, the basic step towards SC development is the education of city citizens. Thus, education must be the primary focus of a SC in order for its residents to thrive. Education inside SCs has not been very well studied yet; therefore, it is possible to advance this sector in numerous areas, such as e-learning, smart dashboards, smart campuses \cite{AliAlnoman}, interactive multimedia \cite{ALmsiedeen12202111}, the metaverse \cite{TarmizanKusuma}, virtual reality \cite{mostafaBourhim}, smart classrooms \cite{JianKang}, and so on.

Furthermore, there is some research done in the area of AI techniques in numerous domains of SC \cite{HerathH, ElGhatiOmar}. SCs are continuously including new services and applications for citizens to use based on AI. AI aims at improving the performance of the city in some domains, such as traffic management and city monitoring. Actually, more studies should be done on AI techniques. AI has not been very well exploited yet, so it is possible to utilize it in several areas of SC, such as cybersecurity, controlling pollution, parking systems, public transportation, waste management, traffic management, energy tracking, security, encryption, and so on.
Moreover, blockchain is a technology for resolving privacy and data management concerns in the context of SCs. Thus, there are some studies to be done on blockchain techniques. The blockchain is the king of transparency.
Several research opportunities for SCs need further study. Figure \ref{fig:researchOpportunities} presents some research opportunities for SCs based on the author's point of view.

In conclusion, several factors are behind the success story of Amman as a SC. Some of those factors are using the right policies and regulations, applying the correct administrative conditions, obtaining the right funds for SC initiatives across different sectors in the city, \etc

\section{Summary} \label{sec:Summary}

In summary, the implementation of SCs is still missing many points: the definition of the SC concept; choosing the appropriate architecture of SC; addressing the open and domain challenges of SC; following a SC roadmap where future phases that the city should take are obviously planned and designed in advance; settling the best practice guidelines in the field of SC; and discussing research opportunities for SC. 

In order to create smart, green, digital, and human-centric cities of the future, there is a need to give more care to education, investments, strategies, and the sharing of experiences. In that scenario, integrating inclusion with digital transformation will still be a big challenge for all types of cities around the world.

In conclusion, this work has analyzed whether Amman has a SC vision, a platform, a designed roadmap, a dedicated department focused on the evolution of Amman as a SC, the SC domains the city covers, the projects that have been executed in Amman, and the technologies used.

\newpage

\bibliographystyle{unsrtnat}
\bibliography{references}

\vfill
\color{blue}\rule{\textwidth}{1.50pt}
\begin{wrapfigure}{l}{25mm} 
	\includegraphics[width=1in,height=1.25in,clip,keepaspectratio]{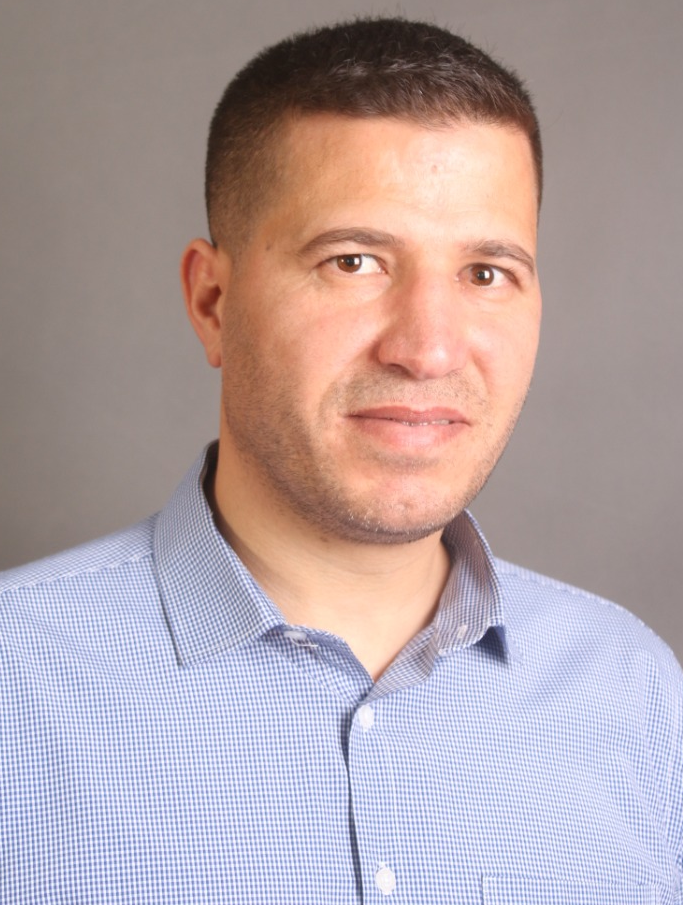}
\end{wrapfigure}\par 
\color{black}
\textbf{Ra'Fat Al-Msie'Deen} is an Associate Professor in the Software Engineering department at Mutah University since 2014. He received his PhD in Software Engineering from the Université de Montpellier, Montpellier - France, in 2014. He received his MSc in Information Technology from the University Utara Malaysia, Kedah - Malaysia, in 2009. He got his BSc in Computer Science from Al-Hussein Bin Talal University, Ma'an - Jordan, in 2007. His research interests include software engineering, requirements engineering, software product line engineering, feature identification, word clouds, and formal concept analysis. Dr. Al-Msie'Deen aimed to utilize his background and skills in the academic and professional fields to enhance students expertise in developing software systems. Contact him at \href{mailto:rafatalmsiedeen@mutah.edu.jo}{\faEnvelope \space }\href{mailto:rafatalmsiedeen@mutah.edu.jo}{ rafatalmsiedeen@mutah.edu.jo}. Also, you can reach him using different alternatives: \href{https://rafat66.github.io/Al-Msie-Deen/}{\faGithub \space }\href{https://rafat66.github.io/Al-Msie-Deen/}{ author's page @ github.io}, \href{https://www.linkedin.com/in/ra-fat-al-msie-deen-08895062/}{\faLinkedin \space }\href{https://www.linkedin.com/in/ra-fat-al-msie-deen-08895062/}{ LinkedIn}, \href{https://www.researchgate.net/profile/Rafat-Al-Msiedeen}{\includegraphics[width=0.022\textwidth]{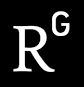}\space}\href{https://www.researchgate.net/profile/Rafat-Al-Msiedeen}{ ResearchGate}, or \href{https://orcid.org/0000-0002-9559-2293}{\includegraphics[width=0.022\textwidth]{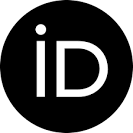}\space}\href{https://orcid.org/0000-0002-9559-2293}{ Orcid}. 
\newline
\color{blue}\rule{\textwidth}{1.50pt}
\newline
~\\
\color{red}\textbf{$\rhd$}\color{black}\textbf{ To cite this version:}

\color{red}\textbf{|}\color{black}\space\space\space~~~~R. Al-Msie'deen, "\color{blue}Amman City, Jordan: Toward a Sustainable City from the Ground Up,\color{black}", ar{X}iv, pp. 1-12, 2024.

\end{document}